# Flexible Thermoelectric Active Cooling Garment to Combat Extreme Heat

*Tianshi Feng[1], Jiedong Wang[1], Ethan Sun[2], Antonio Di Buono[3], Renkun Chen[1,*]*

[1] Department of Mechanical and Aerospace Engineering, University of California San Diego, La Jolla, California 92093-0411, USA

[2] The Bishop's School, 7607 La Jolla Blvd, La Jolla, California 92037-4799, USA

[3] National Nuclear Laboratory, Birchwood Park, Warrington, WA3 6AE, United Kingdom

*E-mail: rkchen@ucsd.edu

**Keywords:** thermoelectric, personal cooling, heat wave, wearable

**Abstract**

With the increasing frequency, intensity, and duration of extreme heat events due to climate change, heat-related diseases or even mortality have become more prevalent. An efficient personal cooling strategy can mitigate heat stress by regulating the skin temperature within the thermal comfort zone. However, lightweight, wearable, and sustainable cooling garments are unavailable today. Here, we developed a TED-based cooling garment and demonstrated its effectiveness in active personal cooling. The garment is shown to maintain the skin temperature within its thermal comfort zone in a hot environment of up to 40 °C under mild forced convection conditions (air flow speed of 2.2 m s$^{-1}$). Furthermore, we demonstrated a portable cooling system with less than 700 grams of total weight, which includes the TED-based garment, a battery pack, and a temperature controller. The system showed long-term cooling on the skin with varying ambient temperatures from 35 to 40 °C. With the advantages of lightweight, flexible, controllable and long-term effective cooling, the TED cooling garments described in this work can contribute to enhanced health and comfort in an increasingly hotter climate.

# 1. Introduction

Global warming has already raised the Earth's average surface temperature by 1.09 °C since the pre-industrial level[1]. The warming has been causing extreme heat events that are increasingly more frequent and intense around the globe[2]. Once the average temperature increase reaches 1.5 °C, an even greater heat wave magnitude is expected over many regions in the world, potentially exposing about 14% of the world population to extreme heat events at least once every five years[3]. Despite public health campaigns to provide resources, emergency department visits, hospitalizations, and deaths from heat stress have been increasing during heat waves, with up to 30% of heat-related deaths globally in the last 30 years[4]. The scenarios became more severe for those regions without enough nighttime cooling or insufficient infrastructure[5]. Even in developed regions such as Europe and North America, insufficient cooling could still lead to unexpected mortality[6]. Therefore, efficient cooling solutions are in high demand to combat extreme heat.

Among different cooling strategies, personal cooling devices and garments, which can maintain the thermal comfort of a person by directly cooling the skin with a wearable system, attract more attention due to their portability and accessibility, especially for those who work outdoors during extreme heat or otherwise lack access to air conditioning[7]. Different personal cooling garments have been developed and reported. They can be summarized in the following three main categories. (1) Garment embedded with ice (e.g., Stacool industrial vest[8]) or phase change materials (PCMs, such as Texas cool vest[9], Techniche phase change vest[10]). They are heavy (2.5-3 kg) and have a short cooling duration as the stored ice or PCMs have limited latent heat. (2) Cooling vest integrated with a portable chiller or refrigerator to continuously make chilled water. The cooling duration is prolonged to 6-8 hours per charge of the batteries for the chiller. However, the refrigerator makes the system very bulky and expensive (e.g., the backpack ice chiller system is weighted at 6 kg and costs $2,500[11]). (3) water evaporative cooling vest (e.g., Techniche evaporative vest[12]). It can provide over 5 hours of cooling due to the large latent heat of water evaporation. However, the vest needs to be soaked with > 4 kg of water to provide a long cooling duration (based on the latent heat of water evaporation, 2200 kJ/kg). In addition, the evaporated water vapor could be in contact with the skin and cause discomfort.

Thermoelectric devices (TEDs) are solid-state refrigeration devices that can pump the heat against a temperature difference, and thus are suitable for personal cooling applications in hot environments. They have attracted significant interest in recent years due to their solid state nature, readily adjustable cooling power and temperature, small form factor, and light weight[13]. There have been several prior works on personal cooling garments using TEDs. However, the challenge lies in dissipating the heat from the hot side of the TEDs to the environment. Xu *et al.* introduced a water circulation system inside a vest integrated with TEDs, which could deliver over 300 W of cooling power in the form of cold water at 30 °C ambient temperature[13b]. Li *et al.* built a TED cooling system using water as a coolant to maintain skin temperature within the skin's thermal comfort zone for up to 1 hr at 34 °C ambient temperature[14]. Zhang *et al.* developed a TED-based active liquid cooling garment in a vehicle to maintain skin temperature within the comfort zone when the cabin temperature was up to 40 °C[15]. These TED garments with a liquid coolant loop have demonstrated sufficient cooling capacities even at high ambient temperatures, but they possess heavy heat sinks for the liquid coolant and are thus bulky. Lou *et al* reported a TED-based system using air as the coolant, which showed a maximum cooling power of 15.5 W at 26.1 °C ambient temperature[16]. The air-based garment is lighter (< 1 kg), but it is not capable of providing sufficient cooling at higher ambient temperatures due to the low thermal conductivity and low heat capacity of air. Additionally, large and rigid TEDs (single device area greater than 16 cm$^2$) were used in these studies, which made the garments not flexible or wearable. Recently, Sony released their latest version personal cooling solution with a relatively small active cooling size (<12x10 cm$^2$)[17]. However, the small local cooling area may not create larger enough cooling sensations for the entire body[18]. The overall device (12.5×2.3×13.7 cm$^2$) is still rigid.

Instead of circulating a coolant (water or air) inside the garments, flexible TEDs can make conformal contacts with the skin to provide direct cooling without bulky heat sinks or a coolant. In our previous work[13a], we developed a flexible TED by replacing rigid ceramic substrates in conventional TEDs with a polymer (Ecoflex). We also optimized the TED design to maintain a skin temperature of 33 °C within the thermal comfort zone at an elevated ambient temperature of up to 36 °C without an external heat sink. Subsequent studies by others have achieved similar or better performance using flexible TEDs by integrating various heat sinks, such as PCMs, porous metal foams, hydrogels, and liquid metals[19]. However, all these prior flexible TED cooling studies are limited to single TEDs or relatively small cooling areas on the skin, such as on the arm. There has been no study showing a larger and wearable active cooling garment working in an ambient temperature resembling extreme heat conditions (e.g., 40 °C).

In this work, we integrated multiple small TEDs into a fabric to make a flexible and wearable active cooling garment, as shown in **Figure 1**. Dyneema fabrics with high thermal conductivity were used to spread the cooling effect between the TEDs. Due to the relatively small size of individual TEDs (2.5×2.5 cm$^2$), the garment is still flexible overall despite the rigidity of individual

TEDs (Figure 1(b)). We specifically designed an active cooling system targeting the human backside to demonstrate its effectiveness in larger functional areas, extending beyond the scope of previous studies. Notably, there is a strong correlation between cooling the back and the overall body's thermal regulation, as compared to cooling other body areas. This targeted cooling approach significantly influences the body's overall thermal sensations, offering a more comprehensive and effective solution[20]. By using the TED design we developed earlier[13a], the garment can maintain the skin temperature within the thermal comfort zone at metabolic heat flux up to 100 W m$^{-2}$ at 32 °C ambient under natural convection conditions on a thermal manikin. By incorporating mild forced convection condition with an air flow speed of 2.2 m s$^{-1}$ provided by a small fan, the garment can provide sufficient cooling to maintain the skin thermal comfort at 40 °C ambient temperature. Testing on humans also shows the capability of maintaining skin thermal comfort in 36 and 40 °C ambient temperatures with the natural and forced convection conditions respectively. With a temperature controller and a battery pack, we further show that the skin temperature can be automatically regulated to the thermal comfort temperature in varying ambient temperatures up to 40 °C. These results suggest that the TED-based flexible cooling garments can be an effective strategy to deliver thermal comfort during extreme heat.

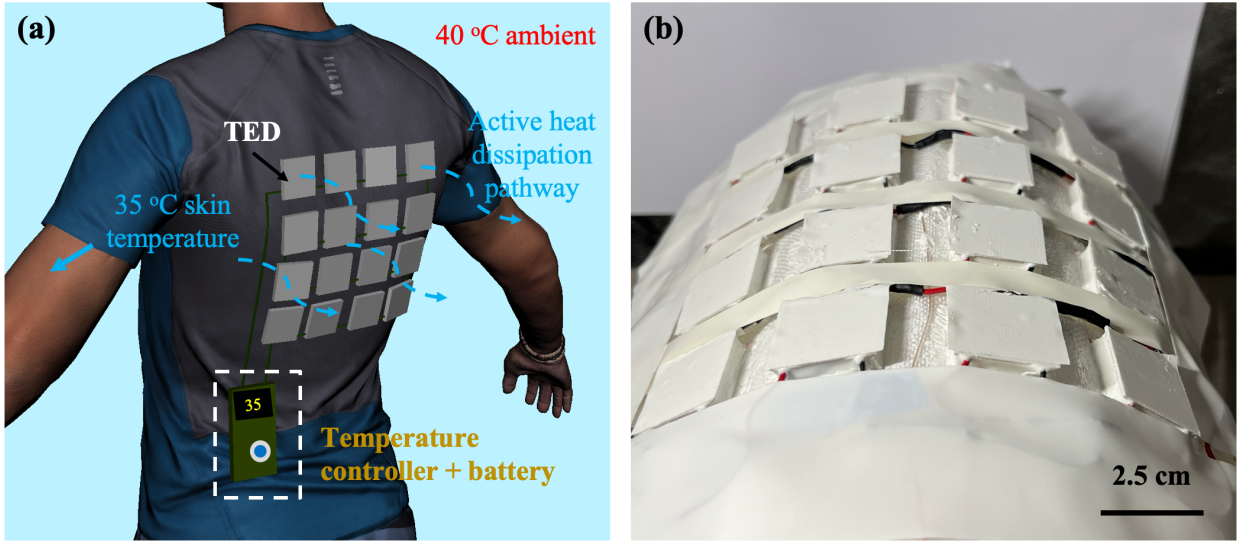

**Figure 1:** Concept of TED-based flexible cooling garment. (a) Schematic of a fabric integrated with multiple small TEDs with efficient heat pumping and dissipation from the skin (~35 °C) to a hot environment (~40 °C). The garment is integrated with a temperature controller and a battery pack. (b) Photograph of a flexible TED-integrated fabric placed on a curved substrate

## 2. Results and Discussion

### 2.1. Characterization of Single TED

The cooling performance of a single TED can be analyzed by considering the Peltier cooling effect on the cold or skin side, parasitic heat conduction from the hot to the cold sides of the TED, joule heating, and heat dissipation from the hot side to the environment [13a]:

$$Q_{skin} = S \cdot I \cdot T_{skin} - G_{TED}(T_H - T_{skin}) - \frac{I^2 R}{2} \qquad (1)$$

$$Q_{hot} = S \cdot I \cdot T_{Hot} - Q_{cond} + \frac{I^2 R}{2} = h_{loss}(T_H - T_{amb}) \qquad (2)$$

where $Q_{skin}$ is the heat flux removed by the TED from the skin, which is also equal to the metabolic heat from the skin, $Q_{hot}$ is the heat dissipation from the hot side of the TED to the air, $S$ is the total

Seebeck coefficient of all the p and n type TE pillars in the TED, $G_{TED}$ is the total thermal conductance of the TED, $R$ is the total resistance of the TED, $I$ is the applied current, $h_{loss}$ is the heat transfer coefficient (HTC) from the hot side of the TED to the air, which includes the contributions from radiation and convection heat transfer, $T_H$ is the temperature of the hot side of the TED, and $T_{skin}$ is the temperature of the cold side of the TED, which would also represent the skin temperature when the device is used for personal cooling, $T_{amb}$ is the ambient temperature. These equations show that as the electrical current applied to the TED increases, the Peltier cooling effect increases linearly, but the Joule heating increases quadratically. Therefore, there is an optimum current to reach the maximum cooling effect and minimum skin temperature for a given set of conditions such as metabolic heat flux, ambient temperature, and the HTC to the air.

We characterized the single TED performance by identifying the optimum current for minimum skin temperature at different ambient temperatures. **Figures 2(a) & 2(b)** show the schematics and photos of the testing setup. The experimental details can be found in the Experimental Section. We define the temperature between the heater and TED as the skin temperature ($T_{skin}$). **Figure 2(c)** shows two infrared (IR) images indicating that the skin side was either heating (left) or cooling (right). When the skin side was cooling, the top surface of the TED was heated up to 54 °C, which enabled the heat dissipation to the 40 °C ambient. We first identified the optimum current to the TED when there was zero heat load from the thin film heater, by measuring the skin temperature versus applied current at different ambient temperatures (21, 29, 33, and 39 °C), as shown in **Figure 2(d).** At different ambient temperatures, the skin temperature first decreased with increasing current due to the Peltier effect. As the current kept increasing, the skin temperature reached the minimum and then increased again due to the Joule heating. The TED reached its minimum skin temperature at a current of around 200-250 mA under the natural convection condition.

A Multiphysics model based on COMSOL was developed to evaluate the single TED performance. The details of the COMSOL model are described in the Experimental Section and in the Supplementary Information (Note S3). The simulation results show good agreement with our experimental data as shown in **Figure 2(d).** The slightly lower skin temperature from the model can be attributed to the simplifications made in the model, e.g., neglecting the thermal and electrical contact resistance.

To quantify the cooling effect the TEDs would induce if they were placed on human skin, a heat load was applied to the device to mimic the metabolic heat. The metabolic rate varies under different conditions. Here, we selected ~80 and ~100 W m$^{-2}$ for an individual engaging with light activities[21] as the representative metabolic rate. The raw data of skin temperature ($T_{skin}$) as a function of the TED current is shown in the Supplementary Figure S4. **Figure 2(e)** summarizes the minimum $T_{skin}$ as a function of ambient temperature ($T_{amb}$) at 0, ~80, and ~100 W m$^{-2}$ heat loads. As ambient temperature increased, the minimum skin temperature increased for all the heat load conditions. When the heat load was applied, the minimum skin temperature also increased at the same ambient temperature. For example, at room temperature, without the heat load, the minimum skin temperature was as low as ~14 °C, while increased to ~23 °C with the 96.4 W m$^{-2}$ heat load. Even with the heat load, the active cooling effect was still strong: the minimum skin temperature can be stabilized at ~37 °C at 39 °C ambient temperature. Therefore, the active cooling performance of a single TED has been demonstrated with a heat load of up to 100 W m$^{-2}$. We note that all skin temperature ($T_{skin}$) measurements were taken at steady state. The skin temperature reached its steady-state value within 5 minutes after the activation of the TED, as shown in Figure S2(b). It remained stable for up to 30 minutes, and in some cases, even longer, as long as the operational conditions (ambient temperature and applied current) remained consistent.

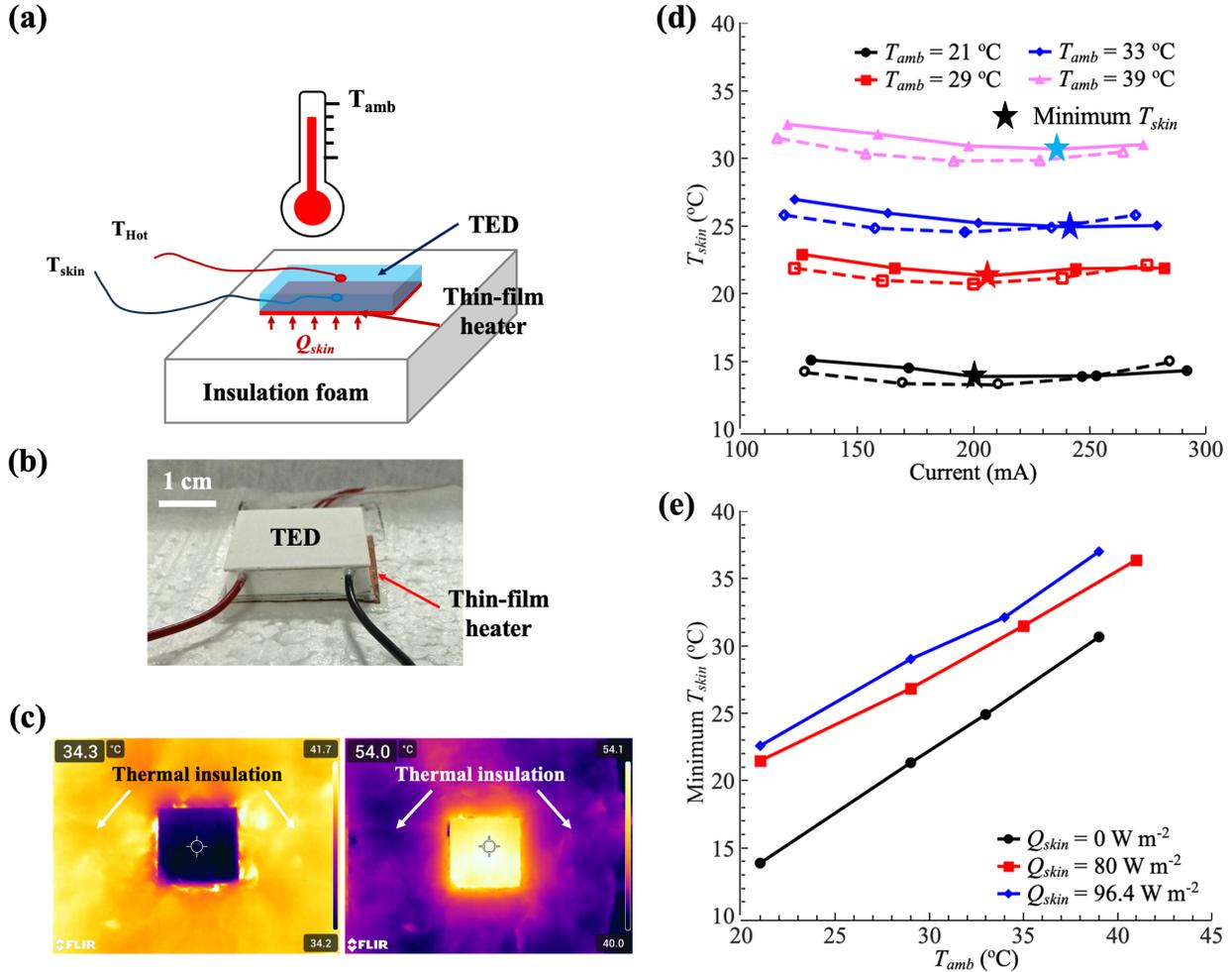

**Figure 2.** Characterization of single TED cooling performance. (a) Schematic and (b) photo of the setup to characterize single TED performance. The test setup was placed in a tent with controllable ambient temperature $T_{amb}$. A thin film heater with controlled heat flux $Q_{skin}$ is placed under the TED. The temperature measured at the heater and underneath the TED is defined as the skin

temperature ($T_{skin}$). The top hot side of the TED is $T_{hot}$. (c) IR images of the TED with the top surface under cooling (left) and heating (right) conditions at 40 °C ambient. (d) The skin temperature as a function of the applied current at different ambient conditions without a heat load; solid lines are from experiments and dashed lines are from the COMSOL model; (e) Minimum skin temperature as a function of ambient temperatures with different heat loads.

*2.2. Development of TED-based cooling garment*

We now examine the performance of the larger active cooling garment consisting of a fabric integrated with 16 small TEDs, as shown in **Figure 1**. To maintain the flexibility of the garment, there needs to be a certain distance between each rigid TED, which requires good heat spreading capability within this region to maintain temperature uniformity. To achieve this, we selected a high thermal conductivity fabric based on Dyneema, an ultra-high molecular weight polyethylene. The reported thermal conductivity of Dyneema ranges from ~14.2 to 28.4 W m$^{-1}$ K$^{-1}$[22], much higher than traditional cotton fabrics (0.026-0.065 W m$^{-1}$ K$^{-1}$)[23]. Therefore, we used a Dyneema composite fabric (Rockywoods Fabrics LLC)[22] in our cooling garment. We measured the temperature in the middle of the fabrics between two adjacent TEDs as shown in **Figure 3**. By incorporating the high thermal conductivity fabric, the middle region can be efficiently cooled compared to a traditional cotton fabric. As shown in the IR images in **Figure 3(b) & (d)**, more uniform and lower temperature was observed in the areas between TEDs in the Dyneema composite fabric compared to the cotton fabric, demonstrating better heat spreading. The same effect is also shown in **Figure 3(e)** displaying the temperatures measured at two different locations: right under a TED (labelled as $T_1$) and in the area between two TEDs ($T_{middle}$). For the cotton fabric, although there was a large cooling effect under the TED, there was nearly no cooling on the fabric in between the TEDs. On the Dyneema fabric, the temperature difference between the two locations was only ~2 °C at the optimum current. As a result, the minimum skin temperature was reduced to ~15 °C in the middle of the Dyneema fabric compared to ~20 °C in the cotton case.

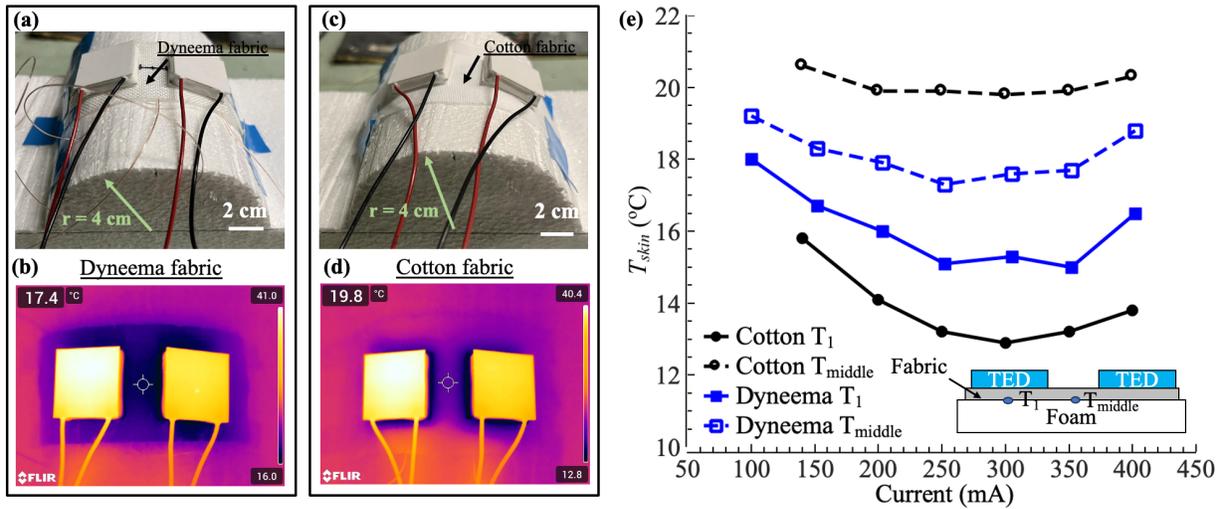

**Figure 3.** Temperature uniformity between Dyneema and cotton fabric. Thermal performance characterization of a Dyneema (a) and cotton (b) based fabric with two TEDs covering an arc-shaped Styrofoam substrate with a radius of 40 mm. Two thermocouples were placed at two different locations on the fabric; the first one ($T_1$) was right under one of the TEDs, and the second was in the middle ($T_{middle}$) between two TEDs. IR image at steady state operation of the TEDs on Dyneema (c) and cotton (d) fabrics. (e) Measured steady state temperature as a function of applied current at two different locations: solid lines represent the thermocouple under a TED ($T_1$) and the dashed lines represent the thermocouple in between the TEDs ($T_{middle}$) The inset shows the schematic of the thermocouple locations.

## 2.3. Performance of TED-based cooling garment on manikin

As shown in **Figure 4(a)**, we tested the cooling garment performance on a manikin. The garment can be worn conformally on the manikin and was further shown good contact on a curved substrate with a radius curvature of 7.5 cm in **Figure 1(b)**, thanks to the flexibility due to the spacing between individual TEDs. The schematic of the experimental setup to measure the cooling performance is illustrated in **Figure 4(b),** with details described in the Experimental Section. We tested the garment with metabolic heat loads of 50, 100, and 150 W m$^{-2}$. The tested metabolic heat loads cover a broad range of human activities, from relaxation to moderate work tasks, including bricklaying, standing while washing dishes, and walking at a rate of 3 km hr$^{-1}$[21a, 24]. Before the current was applied to the TEDs, the skin temperature reached the steady state for at least 10 mins at different operation conditions. Then a current was applied to the TEDs for 20 mins. **Figure 4(c)** shows typical temporal behaviors of the measured skin temperature when the current was applied on the TEDs for two different cases: natural convection at 32 °C ambient temperature with 205 mA current and forced convection at 40 °C ambient temperature with 490 mA current. The skin temperature rapidly cooled down to its minimum value within a few mins. The time for the skin temperature to reach its minimum value at forced convection condition was longer, because of the larger temperature reduction (~47 °C to ~35 °C, compared to from ~39 °C to ~35 °C under the natural convection condition). In both natural and forced convection cases, the skin temperature was stabilized within 15 mins of applied current and maintained for up to 20 mins. Based on the operating principle of the TED, once it reaches a steady state under consistent operating and ambient conditions, its cooling performance is expected to remain stable. The time-dependent temperature response is provided in Supplementary Information, Figure S5. At an elevated ambient temperature of 40°C, the skin temperature fails to stabilize without airflow. However, a steady state is achieved with the presence of air flow, which enhances heat dissipation on the hot side of the TED. Therefore, we extracted the average steady-state skin temperature at about 20 mins after turning on the current in all the subsequent measurements on the manikin.

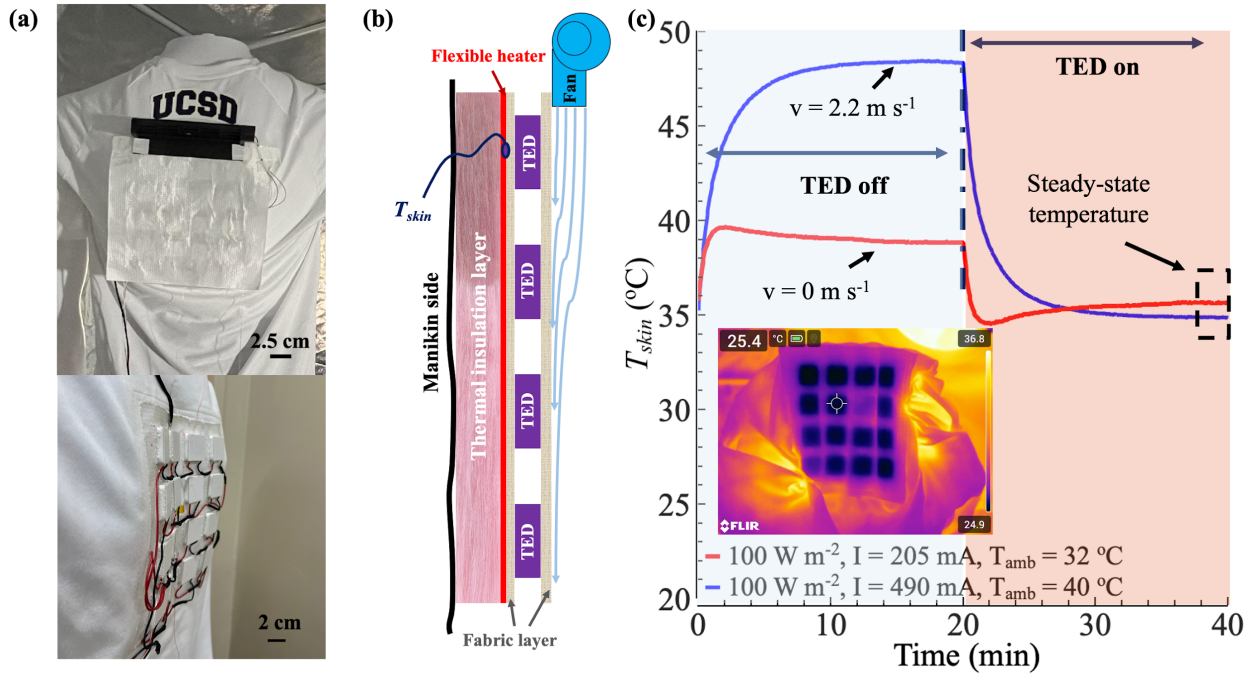

**Figure 4.** (a) Photo of TED-based cooling garment; top: back view; bottom: side view, showing the garment complies with the curvature of the manikin's back. (b) The schematic of the experimental setup for characterization of cooling garment performance on the manikin. Similar to the single TED testing, a thin film heater is placed on a thermal insulation foam, and the TED-based garment is placed on top of the heater. In the forced convection case, a fan is attached to the top of the garment that blows air downward over the garment. The skin temperature is measured by inserting a thermal couple between the fabric and the heater underneath the TED; (c) The real-time temperature response at different operation conditions with heat load at 100 W m$^{-2}$. Blue curve: forced convection with 2.2 m s$^{-1}$ wind speed at ambient temperature of 40 °C; Red curve:

natural convection at ambient temperature of 32 °C. The inset IR image shows the steady state cooling effect from the cooling garment under zero heat load at 32 °C ambient temperature.

We first tested the TED cooling garment without a fan (i.e., natural convection) under three different heat loads (50, 100, and 150 W m$^{-2}$). The measurements on the garment were similar to the ones done on single TEDs to identify the optimum operation current under different ambient temperatures and heat loads. The results are shown in **Figure 5(a)**. At these two heat loads, the skin temperature reached its minimum value around 200 mA which is consistent with the single TED measurements shown in **Figure 2(b)**. To maintain the skin temperature within the skin thermal comfort zone (33.8-35.8 °C)[20b], the maximum operation ambient temperatures were identified as 32 and 35 °C for the 100 and 50 W m$^{-2}$ heat load respectively. There was no sufficient cooling effect for the 150 W m$^{-2}$ heat load when the ambient temperature was above 30 °C.

To further improve the TEC cooling garment performance, better heat dissipation from the hot side to the environment is needed, e.g., by enhancing the heat transfer coefficient (HTC, the $h_{loss}$ term in Equation 2). A higher HTC would enable the heat removal with a smaller temperature difference between the hot side and the environment, or greater heat dissipation for the same temperature difference. We used a small fan to induce a mild wind with a speed of ~2.2 m s$^{-1}$ flowing over the exterior surface (i.e., the hot side) of the TEDs, thereby creating a forced convection HTC of ~37 W m$^{-2}$ K$^{-1}$, which is about 6 times higher than the natural convection HTC. Combined with the radiation HTC of ~6 W m$^{-2}$ K$^{-1}$, this leads to a total HTC of ~43 W m$^{-2}$ K$^{-1}$. The estimation of the heat transfer coefficient can be found in the Supplementary Information (Note S2). The effect of the enhanced HTC by the forced convection was clearly shown in **Figure**

**5(e) and (f)**: under the same current, the TED hot side temperature was greatly reduced compared to the natural convection case (54.4 vs 75.6 °C). Consequently, a lower skin temperature was achieved due to the reduced parasitic heat leakage from the hot to cold sides of the TEC (the $G_{TED}(T_H - T_{skin})$ term in Equation 1). As shown in **Figure 5(d)**, the optimum operation current to reach the minimum skin temperature was increased with the forced convection, which resulted in larger Peltier cooling (the $ST_cI$ term in Equation 1). The enhanced HTC with the forced convection allows the dissipation of the larger amount of joule heating due to the higher current.

**Figure 5(e)** summarizes the measured minimum skin temperatures under various applied currents for the three heat loads (50, 100, and 150 W m$^{-2}$) as a function of ambient temperature. As a reference, we also did a control experiment without the TEDs, i.e., the same flexible heater was covered with a conventional T-shirt (**Figure 5(b)**). The thermal comfort zone of the human back skin (33.8-35.8 °C)[20b] is shown as the shaded area in the figure. We can define the maximum operation ambient temperature when the skin temperature reaches the upper bound of the skin comfort zone (35.8 °C). **Figure 5(e)** shows that the maximum operation ambient temperatures are 37 and 43 °C, respectively, under the natural and forced convection conditions respectively when the heat load was 50 W m$^{-2}$, and are 32 and 40 °C, respectively, under the natural and forced convection conditions respectively when the heat load was 100 W m$^{-2}$, and are 27 and 37 °C, respectively, under the natural and forced convection conditions respectively when the heat load was 150 W m$^{-2}$. On the other hand, for the control experiment without the TEDs, the skin temperature reached the maximum thermal comfort temperature when the ambient temperature was only about 25 °C, which is consistent with our experience of comfortable ambient temperature without any external thermoregulation[25]. Therefore, the results from **Figure 5(e)** clearly demonstrate the effectiveness of the TED garments to actively cool the skin in hot environments, up to 32 and 40 °C ambient temperature with natural and forced convection conditions, respectively, for metabolic heat of 100 W m$^{-2}$. The maximum operational ambient temperature slightly decreased to 37°C for an even higher metabolic heat rate of 150 W/m². We further evaluated the impact of different thermoelectric (TE) materials with varying ZT values, ranging from 0.71 (in this work) to 1.48 [26] through simulation. The summarized results in Supplementary Information, Figure S6(b), show a >40% enhancement in coefficient of performance (COP), approximately 60% reduction in TE material weight, and around 40% reduction in TE pillar height when replacing the current TE material with MgAgSb-based material. We believe that a more lightweight, compact, and high-performance TED-based cooling garment can be achieved.

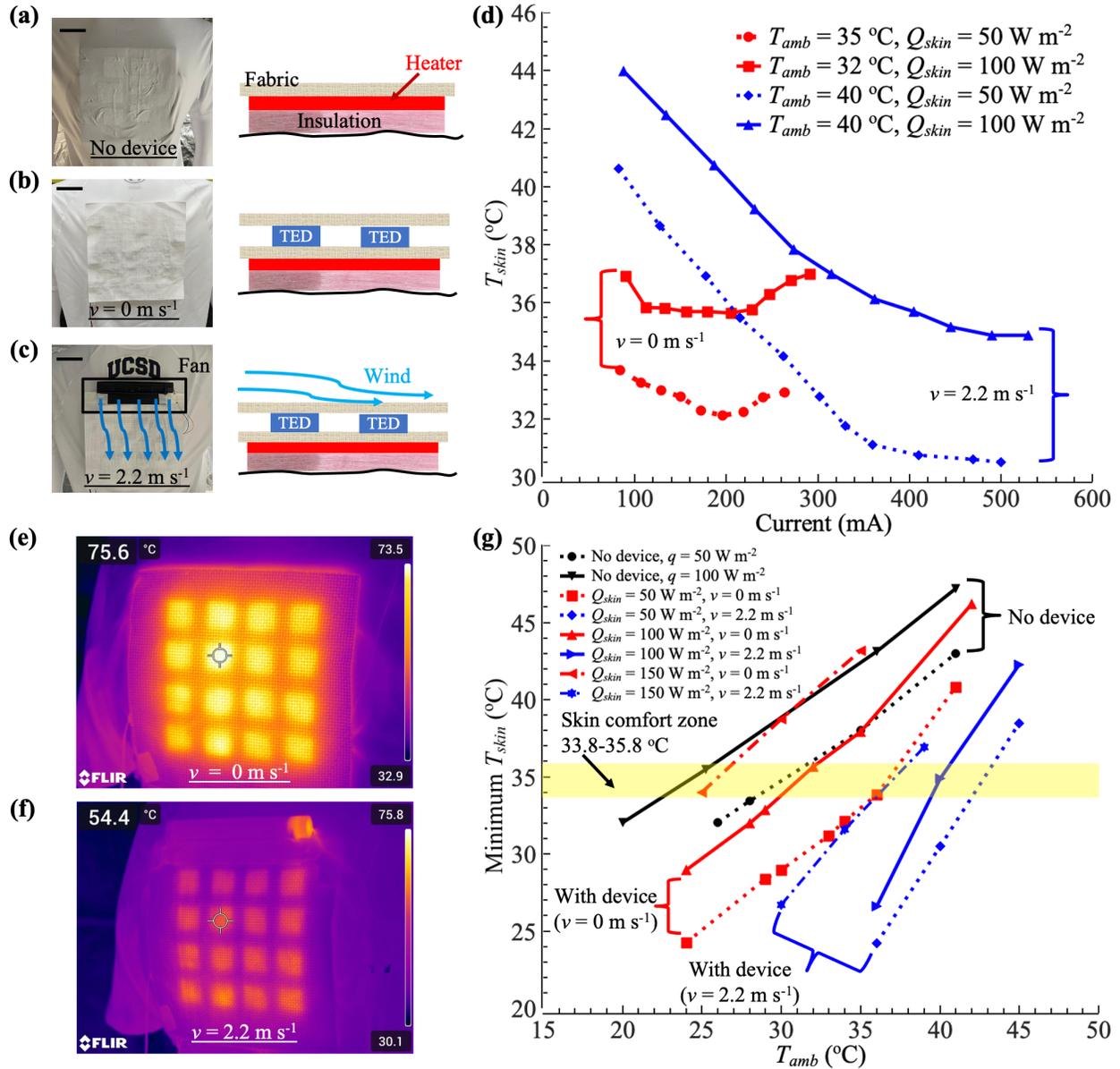

**Figure 5:** TED cooling garment performance on manikin model under different operation conditions, blue represents forced convection at mild windy conditions with wind speed 2.2 m s$^{-1}$; red represents natural convection; and black represents non-device conditions: (a-c): The photo and schematic of different test conditions, the scale bar is 5 cm; (d) The skin temperature as a function of applied current at different ambient temperature and heat loads; (e) & (d): IR images from the steady-state hot side temperature with or without wind respectively; (g) The minimum skin temperature as a function of ambient temperature at different operation conditions.

At this stage, we have successfully demonstrated the effectiveness of the TED-based cooling garment in actively cooling the skin under varying ambient conditions and metabolic heat loads at steady state. However, for real-world applications, long-term stability and continuous thermal cycling are essential. The real-time temperature response is shown in Supplementary Information, Figure S5. Within the maximum operational temperature range, the skin temperature stabilized for up to 20 minutes, maintaining a consistent temperature under the same operating conditions, as seen in Figures S5(a) and (d).

To further assess long-term and cyclic stability, we evaluated the device's active cooling performance over 1-hour cycles. As summarized in Figure S5(c), the device was alternately switched on and off, with the skin temperature cycling between low and high states. Each cycle lasted 60 minutes, and a total of six cycles were tested, covering a cumulative duration of over 750 minutes (>12.5 hours). During each cycle, when the TED was activated, the skin temperature stabilized for up to 60 minutes, further confirming its long-term stability. The active cooling effect during each cycle, along with the consistent performance of the device when switched on, demonstrates superior cyclic stability.

### 2.4. Performance of TED-based cooling garment on user

We then demonstrated the performance of the TED garment when worn by an individual. We integrated a temperature controller to automatically control the skin temperature within the thermal comfort zone when the ambient temperature was changing. A battery pack was used as the power supply. The integrated portable TED cooling garment is shown in **Figure 6(a).** The entire device weighs about 700 g including all the components shown in the figure (temperature controller, battery pack, TED garment). The relatively light weight ensures the wearability and portability of the TED garment.

We first quantify the minimum skin temperature achievable by the cooling garment with varying ambient temperature without using the temperature controller. The test individual wearing the TED garment was seated inside a tent with a controlled ambient temperature. At each ambient temperature, the current to the TEDs were varied and the back skin temperature was measured using a thermal couple until the minimum skin temperature was identified, similar to the experiments on the manikin (**Figure 5(d)**). The measured minimum skin temperature at different ambient temperatures is shown in **Figure 6(c).** Without the TEDs, the skin temperature shows a non-linear increase with the ambient temperature, due to the self-thermoregulation of the skin[20b, 27]. Under the natural convection condition, the minimum skin temperature can be maintained within the thermal comfort zone with an ambient temperature of up to 37 °C. By applying the same air flow at 2.2 m s$^{-1}$ with the fan, the ambient temperature could be increased up to 40 °C while the minimum skin temperature was well within the comfort zone. These results are similar to those on the manikin shown in **Figure 5(g)**. The active cooling of the garment was also evidenced from the IR image shown in **Figure 6(b):** during the operation of the TEDs, the hot side temperature increased to above the ambient temperature for heat dissipation to the environment while and other side was cooled down.

We then show the dynamic thermal response of the skin to a changing ambient temperature when the temperature controller was used. The target skin temperature was set to 35 °C in the temperature controller. As shown in **Figure 6(d),** the skin temperature was automatically regulated within the thermal comfort zone by the controller when the ambient temperature changed. At 35 °C ambient temperature, when the current to the TEDs was turned on, the controller adjusted the output power to maintain a nearly constant skin temperature at 35 °C target value. The skin temperature rapidly stabilized at the setpoint within a few mins. We then switched off the current and increased the ambient temperature from 35 to 40 °C. The skin temperature increased above the skin comfort zone due to insufficient cooling. Once the current to the TEDs was switched on and the temperature controller was operational again, the skin temperature went back down to the thermal comfort zone within 3 mins. Moreover, the skin temperature can be maintained within the comfort zone for a long time at 40 °C ambient temperature. **Figure 6(d)** also shows the recorded real-time power consumption in the battery pack. The maximum power consumption was about 16 W when the ambient temperature was 40 °C. Therefore, we can estimate that a 0.5 kg battery pack with an energy density of 250 Wh kg$^{-1}$ can enable at least 7.5 hr of operation at 40 °C ambient temperature. This result shows the feasibility of dynamic thermoregulation to achieve long-time thermal comfort using the TED cooling garment in hot environments.

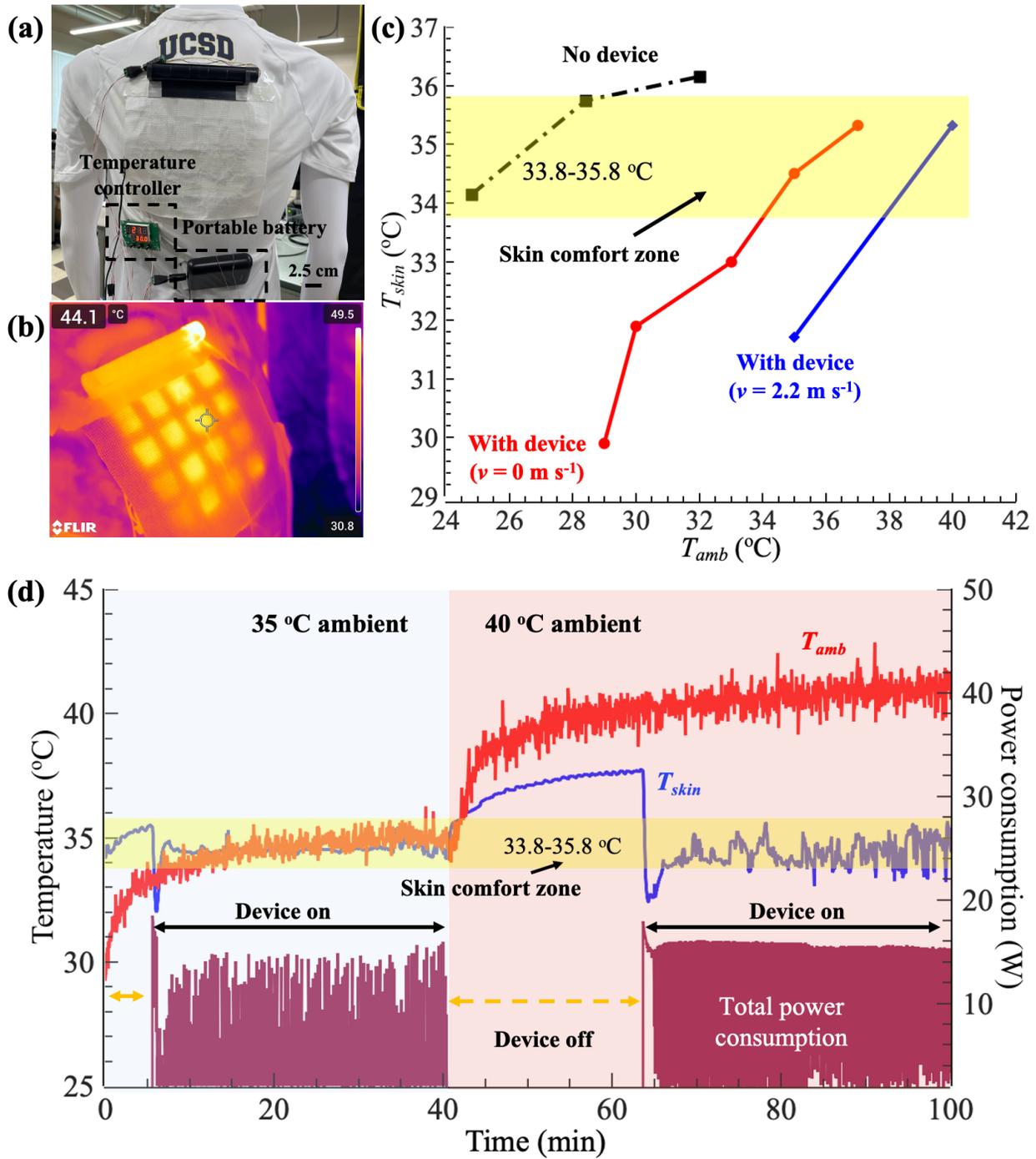

**Figure 6.** TED cooling garment performance on the human body. (a) The prototype of a portable TED cooling garment by integrating a temperature controller and a battery. (b) IR image of the hot side of the TED cooling garment under forced convection condition. (c) The minimum back skin temperature at different ambient temperatures; blue curve represents forced convection wind speed

2.2 m s$^{-1}$; red curve represents natural convection; and black represents the case without TEDs (i.e., only wearing a regular cotton T-shirt. (d) Dynamic thermal regulation on the skin with the TED cooling garment under TED on and off conditions when the ambient temperature changed from 35 to 40 ºC. The setpoint (target temperature) of the temperature controller was set at 35 ºC when the TED was turned on. The red curve represents the real-time ambient temperature, and the blue curve represents skin temperature. The real-time total power consumption is shown in the purple curve (right y-axis).

## 2.5. Safety and Washability

As shown in Figure 5(d), the maximum cooling power was achieved at a current level exceeding 400 mA due to enhanced heat dissipation under windy conditions. This current level appears to exceed the typical safety range (<30 mA) However, it is important to note that this high current only flows through the device and not through the human body in the event of an unexpected leakage. The entire system is powered by a 12 V DC battery, which is considered safe for human contact. According to general safety guidelines, DC voltages up to 30 V are safe under wet skin conditions, and up to 60 V under dry skin conditions[29]. Specifically, previous research has shown that DC voltages up to 24 V are very safe for the human body [30].

Assuming a contact area of 1 mm² and a skin thickness of 1 mm, the skin resistance would be approximately 1 kΩ, given the poor electrical conductivity of human skin (approximately 1 S/m) [31]. Under a 12 V voltage, the resulting current would be around 12 mA, which is well within the safe range. Previous studies have found that a maximum current of 20 mA is safe and tolerable for electrotactile sensations [32]. Therefore, although the possibility of unexpected leakage exists, which is unlikely due to the well-insulated active electrical components, any current that might flow through the skin would remain within the safe limit.

In real-world clothing applications, washability is a crucial consideration. Our cooling garment comprises four active components: TEDs, a temperature controller, a battery, and a fan, as shown in Figure 6(a). The fan, controller, and battery, which are not waterproof, can be easily detached from the garment for washing. The TEDs, however, are protected by waterproof seals between the two alumina substrates, as shown in Figure 2(b), ensuring that no electrical connection pads are exposed to water. Additionally, a waterproof silicone-based thermal adhesive is used to attach the TEDs to the Dyneema fabric, and all wiring connections are embedded within waterproof rubber seals. As a result, the garment can be washed with water after removing the non-waterproof components.

However, the current prototype is not designed to withstand washing with detergents. To improve durability, solvent- and detergent-resistant adhesive sealants could replace the current silicone-based glue, strengthening the seal between the TEDs and the fabric. Companies such as MASTERBOND®, Henkel, and MG Chemicals offer a variety of chemical-resistant adhesive options. We believe that with the adoption of such adhesives, future commercialization of TED-based cooling garments could enhance washability while maintaining their performance.

### *2.6. Future Perspective on Active Cooling Garment with Critical Active Area*

In this study, we designed, fabricated, and evaluated the performance of a TED-based flexible cooling garment using high thermal conductivity Dyneema fabric, covering a total area of approximately 20×20 cm². Our results indicate that Dyneema fabric provides a more uniform cooling effect compared to regular cotton fabric. However, the minimum temperature beneath the TED was slightly higher with Dyneema, due to its larger effective cooling area, as shown in Figure 3. This design allows the back skin temperature to remain within the thermal comfort zone under various operational conditions, which we used as a key metric to evaluate overall device performance.

Localized cooling has a significant impact on whole-body thermal sensations, as part of the broader human thermal regulation system [20c, 33]. However, the critical size of the cooling area and the required temperature drop for substantial effects on overall thermal sensations are not yet well understood. Previous studies suggest that small cooling areas, such as 6.25 cm², only produce a slight reduction in overall thermal sensations [34], while larger cooling areas, ranging from 122 cm² to 384 cm², can significantly reduce sweat rates and enhance the sensation of cold in hot environments[18].

Interestingly, although regular cotton fabric is less thermally conductive, its smaller effective cooling area provides stronger localized cooling. This suggests that it may still offer significant cooling effects under certain conditions. These findings underscore the need for more extensive human subject testing to identify the optimal cooling area and intensity for achieving whole-body thermal comfort.

We believe that our TED-based cooling garment prototype presents a promising solution for mitigating heat stress in hot environments. Future optimization of its size and cooling capacity, informed by further studies, could enhance its performance and broaden its application in real-world scenarios.

## 3. Conclusion

In summary, we successfully fabricated a portable active cooling garment by integrating 16 small individual TEDs onto a high thermal conductivity Dyneema fabric with a total area ~400 cm$^2$. By spacing the individual rigid TED, the garment exhibits good flexibility. The garment cooling performance was demonstrated by rapid cooling (within 3 minutes) of the skin when the TEDs were turned on in a hot environment. Under the natural convection condition, thermal comfort on the skin can be attained with the ambient temperature of up to 35 $^{\circ}$C under the heat load of 100 W m$^{-2}$. With forced convection induced by an air flow at 2.2 m s$^{-1}$ driven by a small fan, the maximum ambient temperature for skin thermal comforts increased to 40 $^{\circ}$C under the same heat load, while the maximum ambient temperature slightly decreased to 37 $^{\circ}$C under 150 W m$^{-2}$ heat load. The enhanced cooling performance with forced convection condition was due to the better heat dissipation at the hot side of TED to the environment. The performance of a complete active cooling garment with a battery pack and a temperature controller was demonstrated on a user in a hot environment up to 40 $^{\circ}$C with the mild forced convection condition. The total weight of the complete garment, including the battery and the temperature controller, was less than 700 g. Based on the measured maximum power consumption of 15 W, we expect over 7.5 hr continued cooling with a ~500-gram battery. This TED-based active cooling garment has the advantages of being lightweight, flexible, and most importantly, controllable and long-term cooling in extreme heat up to 40 $^{\circ}$C. We believe that future TED cooling garments developed based on this strategy can contribute to enhanced health and comfort in an increasingly hotter environment. A simulation-based prediction suggests that high-performance TE materials could enhance garment cooling performance while enabling the device to be lighter and more compact. The technology would be especially beneficial for populations who are more vulnerable to extreme heat, such as outdoor workers (e.g., agricultural, construction, transportation, law enforcement.) and individuals with underlying health conditions.

## 4. Experimental Section

*Characterization of single TED cooling performance:* The individual TEDs were fabricated based on our design reported earlier[13a], which features tall pillars (4 mm) with large spacing between them (3 mm). This design led to a relatively low thermal conductance of the TED, such that the heat from the hot side would not parasitically leak to the cold or skin side. The overall dimension of each TED is 25×25×5.5 mm$^3$. The weight of a single TED is 5.5 g. The test setup for the individual TEDs and the garment is a temperature-controlled chamber, and the ambient temperatures were using a thermocouple. As shown in Figure 2(a), a flexible heater (McMASTER-CARR, 35475K232) with the same size as a TED was sandwiched between the TED and an insulating Styrofoam substrate. Different heat fluxes (e.g., 50 and 100 W m$^{-2}$) were applied to the heater to mimic the human metabolic heating rate. A thermocouple was inserted between the heater and the TED for measuring the temperature, which was referred to as the "skin temperature" because the TED would be in contact with the skin in the real-world application. Another thermal couple was attached to the top surface of the TED to measure the hot side temperature. The TED was powered with a variable DC power supply programmed by a LabView program. At each current level, the operation time of TED is 20 mins to ensure a steady state for extracting the temperatures. Between two consecutive measurements, the TED was switched off for at least 20 minutes to remove any residual heating or cooling effect. The steady state skin temperature is shown in Figure 4(c).

*COMSOL model of individual TED:* The thermal performance of a single TED was simulated via COMSOL Multiphysics. By combining the thermoelectric effect and electromagnetic heating modules in the COMSOL, the cooling effect can be calculated. A stationary study was specifically defined as referring to the experimental steady state. The Seebeck coefficient was set as 196 $\mu$V K$^{-1}$ based on our previous measurement[13a]. The model geometry was built based on the size of the device and the constituent pillars. For a typical device, there are 36 TE pillars, along with the entrapped air, between two alumina substrates. Beneath the bottom alumina substrate, an inward heat flux boundary condition was used as the heat load. On the top alumina substrate, the convection boundary condition was applied with the estimated heat transfer coefficient at each operation condition (more details can be found in the Supplementary Information Note S3). To simulate the thermoelectric effect, the electric potential and ground boundary conditions were introduced for electric current passing through the TED. By changing the potential value, different current or operation conditions could be obtained. Finally, the cold side skin temperature was extracted by averaging the outer surface of the bottom alumina substrate, where the inward heat flux was applied.

*Characterization of TED garment cooling performance on the manikin model:* We integrated a Dyneema composite fabric into the back region of a T-shirt as shown in Figure 4(a). The Dyneema fabric area was around 20×20 cm$^2$, which would be in direct contact with the back skin as an active cooling region. Then, 16 identical individual TEDs were attached to the Dyneema fabrics using a double-side thermal paste with the edge sealing of a thermal glue. The spacing between the TEDs is 1.8-2 cm. The schematic of the cooling garment with the individual TEDs and the corresponding wiring connection is shown in Supplementary Figure. S1. The heater with the same 400 cm$^2$ area was attached on the other side of the fabric to mimic the human metabolic heating. An additional thermal insulation layer was applied beneath the heater to avoid heat loss during the thermal mannikin test (but it was absent on human test). Next, another Dyneema composite fabric was attached to the outer surface of the TEDs to cover the device and wiring. A small 12 V DC fan (OLC Inc. CFR5B12X) was attached to the outside layer to create air flow and the forced convection condition. The fabricated TED cloth on the manikin model is shown in Figure 4(a). The schematic of the cooling test setup is shown in Figure 4(b). The skin temperature was measured using a thermocouple applied between the fabrics and the heater. The manikin model was placed inside a temperature-controlled chamber. A LabView-controlled DC power supply was connected to the TED cloth. The test method was the same as the single TED performance characterization.

*Characterization of TED garment cooling performance on human:* We conducted a cooling performance evaluation on a person wearing the cooling garment. The skin temperature on the person's back that was in direct contact with the TED garment was measured when the person was seated inside the temperature-controlled chamber while wearing a regular T-shirt or the cooling garment. Steady state skin temperature was recorded when the thermocouple reading was stable, typically 20 mins after the chamber temperature was changed. Two types of tests were conducted. In the first test, at each ambient temperature, different current levels from a DC power supply were applied to the TEDs, and the minimum steady state skin temperature was recorded, similar to the tests on the manikin. In the second test, a temperature controller (weight at ~50 g) and a portable battery pack (~268 g) were integrated into the garment, as shown in Figure 6(c). The total weight of the entire TED cooling cloth was measured to be less than 700 g. A setpoint of 35 °C in the temperature controller is the target thermal comfort skin temperature. The controller can regulate the skin temperature, reaching the target temperature by automatically adjusting the output current to the TEDs using the PID control. We also measured the total output power consumption (to the TEDs and the fan) in real time. Two different ambient temperatures were tested: 35 °C and 40 °C. The tested individual wearing the portable TED garment (including the battery pack, temperature controller, and fan) was first sitting in the 35 °C chamber for about 5 mins with the TEDs off, and then for about 35 mins when the TEDs were turned on. Then the ambient temperature was changed to 40 °C while the TEDs were off for ~25 mins and on for ~35 mins. The back skin temperature and the total power consumption were recorded as a function of time throughout the process.

*Ethics Information:* All the experimental procedures on healthy human subjects were conducted following a protocol approved by the Institutional Review Board at University of California San Diego (project no. 130003). Informed written consent from the users was obtained prior to the experiments. By following the protocol, the power applied to the TEDs in the garments was controlled using a temperature controller in order to maintain the skin temperature within the neutral temperature zone. The TEDs are ensured to be electrically insulated from the skin with the dielectric layer in the TEDs as well as the fabric layer between the TEDs and the skin.

**Acknowledgement**

This work is supported by a senate grant from the University of California San Diego (No. RG113937) for material and device fabrication, modeling, and device characterization. The work utilized the TED characterization setup that was supported by a grant from the United Kingdom National Nuclear Laboratory.

**Data Availability Statement**

The data that support the findings of this study are available from the corresponding author upon reasonable request.

# Flexible Thermoelectric Active Cooling Garment to Combat Extreme Heat

Supplementary Information

**Note S1. Schematic of TED-based Cooling Garment**

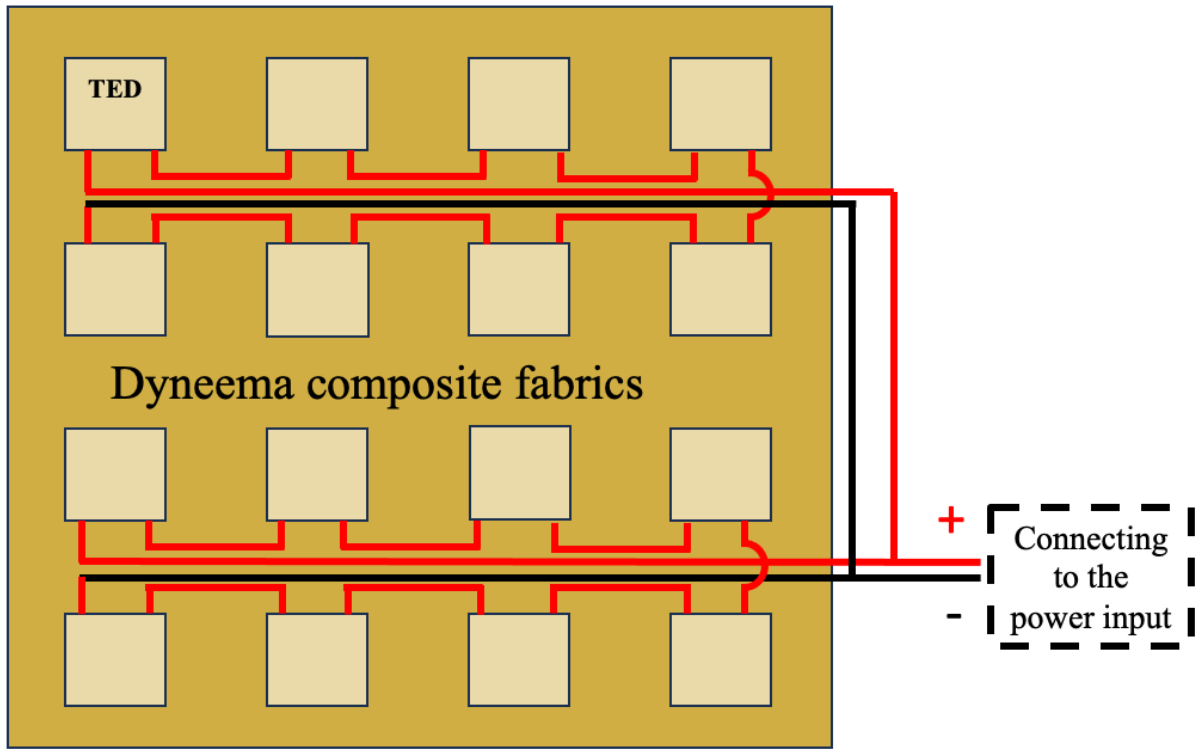

**Figure S1.** Schematic of wire connection of 16 individual TEDs on a Dyneema composite fabrics. The 16 TEDs are divided into two groups. Within each group, the 8 TEDs are connected in series. The two groups of TEDs are connected in parallel to a power source.

**Note S2. Estimation of Heat Transfer Coefficient at the Hot Side of TED**

*Note S2.1. Natural Convection*

By considering the hot side of TED as an isothermal hot surface as shown in **Figure 2 (a)-(c)**, the natural convection heat transfer coefficient can be estimated from the Rayleigh number ($Ra$) and the Nusselt number ($Nu$). In our case, the hot surface is always at the top of the TED. Therefore, the Nusselt number can be described as a function of the Raleigh number as follows:

$$Nu = \begin{cases} 0.54 Ra^{1/4}, & 10^4 < Ra < 10^7 \\ 0.15 Ra^{1/3}, & 10^7 < Ra < 10^{11} \end{cases} \quad \text{(SI-1)}$$

where the Rayleigh number is defined as the product of the Grashof ($Gr$) and Prandtl ($Pr$) numbers as follows:

$$Ra = GrPr = \frac{g\beta(T_s - T_\infty)l^3}{v^2} Pr \quad \text{(SI-2)}$$

where $T_s$ is the TED hot side temperature and $T_\infty$ is the ambient temperature, $\beta$ is the coefficient of thermal expansion. By considering air as an ideal gas, $\beta$ is defined as $\beta = 1/T_\infty$. $v$ is the air kinematic viscosity; $g$ is the gravitational constant; $l$ is the characteristic length which is defined as the perimeter ($P$) divided by the surface area ($A$) as $l = A/P$. The convection heat transfer coefficient can be estimated from the Nusselt number:

$$h = \frac{Nuk}{l} \quad \text{(SI-3)}$$

where $k$ is the air thermal conductivity. By combining the Equation (SI-1)-(SI-3), the natural convection heat transfer coefficient was estimated as a function of the temperature difference between the hot surface and the environment, $T_s - T_\infty$, in Figure S2(a). We further plotted the measured temperature at both cold ($T_C$) and hot ($T_H$) sides of TED at 21 °C ambient temperature in Figure S2(b). In Figure 2 (c), the hot side temperature is 54 °C at the ambient temperature of 40 °C. Therefore, we estimated the natural convection heat transfer coefficient with the temperature difference is 14-15 °C. As a result, the estimated heat transfer coefficient ($h_{conv}$) is about 9.5 W m$^{-2}$ K$^{-1}$.

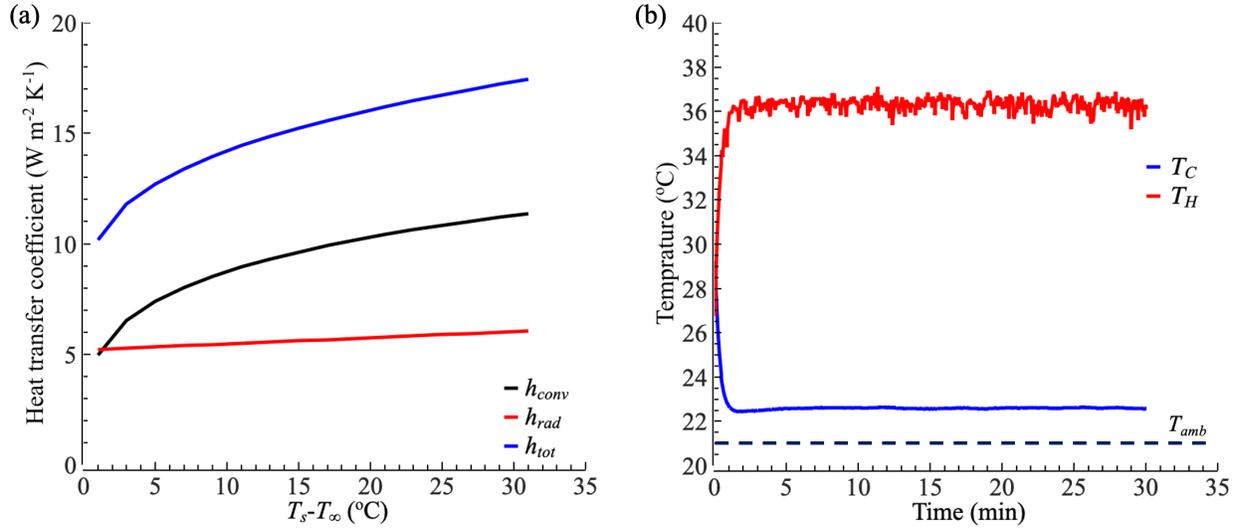

**Figure S2.** (a) Estimated heat transfer coefficient at 21 °C ambient temperature of both natural convection ($h_{conv}$) and radiation ($h_{rad}$); (b) Single TED cold ($T_C$) and hot ($T_H$) side temperature as a function of time with 96.4 W m$^{-2}$ heat load at 21 °C ambient temperature

*Note S2.2. Radiation Heat Transfer Coefficient*

In addition to the natural convection heat loss, the radiation heat loss to the environment is another heat dissipation path in our system as shown in Figure 2(a) & (b) due to the high emissivity of alumina substrates ($\varepsilon \sim 0.9$). Herein, we also theoretically estimated the radiation heat transfer coefficient as the following:

$$h_{rad} = \varepsilon\sigma(T_s + T_\infty)(T_s^2 + T_\infty^2) \qquad (SI\text{-}4)$$

where $\sigma$ is the Stefan-Boltzmann constant ($\sigma = 5.67 \times 10^{-8}$ W m$^{-2}$ K$^{-4}$). The estimated radiation heat transfer coefficient was plotted in Figure S2(a). Assuming ~14-15 °C temperature difference ($T_s - T_\infty$), the radiation heat transfer coefficient is about 5.5 W m$^{-2}$ K$^{-1}$. Therefore, the total heat transfer coefficient of the heat dissipation from the TED hot side to the environment is the sum of the natural convection and radiation heat transfer coefficient which is about 15 W m$^{-2}$ K$^{-1}$.

*Note S2.3. Forced Convection Heat Transfer Coefficient*

As mentioned in the main manuscript, a significant improvement in heat dissipation was observed by introducing a mild windy condition at $v = 2.2$ m s$^{-1}$. To fully understand the mechanism of the improved performance, we theoretically estimated and compared the heat transfer coefficient between natural and forced convection. For simplicity, we considered our system as a flat surface where the bulk air motion is above the hot surface. For a laminar flow, tThe averaged Nusselt number (Nu) and heat transfer coefficient (h) are defined as follows:

$$Nu_L = \frac{hL}{k} = 0.664 Re_L^{1/2} Pr^{1/3}, Pr \geq 0.6 \tag{SI-5}$$

$$Re_L = \frac{\rho v L}{\mu} \tag{SI-6}$$

where $L$ is the length of the TED, $v$ is the wind speed, $\rho$ is the air density, and $\mu$ is the air dynamic viscosity. By assuming a 15 °C temperature difference between the TED hot surface and the ambient, we summarized the convection heat transfer coefficient as a function of wind speed in Figure S3. At 2.2 m s$^{-1}$ wind speed, the convection heat transfer coefficient is estimated to be ~ 37 W m$^{-2}$ K$^{-1}$ which is nearly four times larger than natural convection ~9.5 W m$^{-2}$ K$^{-1}$.

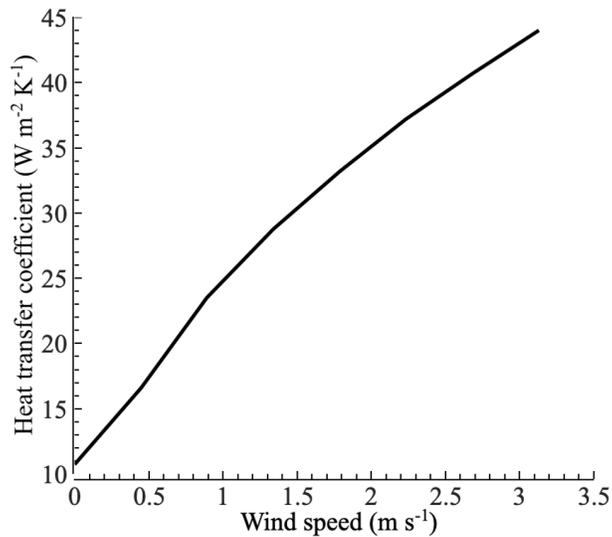

**Figure S3.** The estimated convection heat transfer coefficient as a function of wind speed (note: at zero wind speed, the heat transfer coefficient was estimated based on natural convection)

**Note S3. COMSOL Model of Single TED Performance**

The thermal performance of a single TED was simulated in COMSOL Multiphysics 5.5. The model geometry was built based on the size of the device and the constituent pillars. For a typical device, there are 36 TE pillars with a 1 × 1 mm² cross-section area and a height of 4 mm, along with the entrapped air, between two alumina substrates which is 25 × 25 mm² in size and has a thickness of 0.69 mm. The spacing between each pillar was 3 mm. Beneath the bottom alumina substrates, the inward heat flux boundary was applied to simulate the metabolic heat from the skin. On the top alumina substrate, the convection boundary condition was introduced with the estimated total heat loss coefficient from SI-2 which is ~15 W m⁻² K⁻¹ for natural convection and radiation heat transfer. To verify the model input parameters, we compared the electrical resistance from the model and the real device. The total electrical resistance of the simulated device is ~2.35 Ω, which is close to the real device which has a ~2.37 Ω total resistance.

**Note S4. Single TED Performance Characterization**

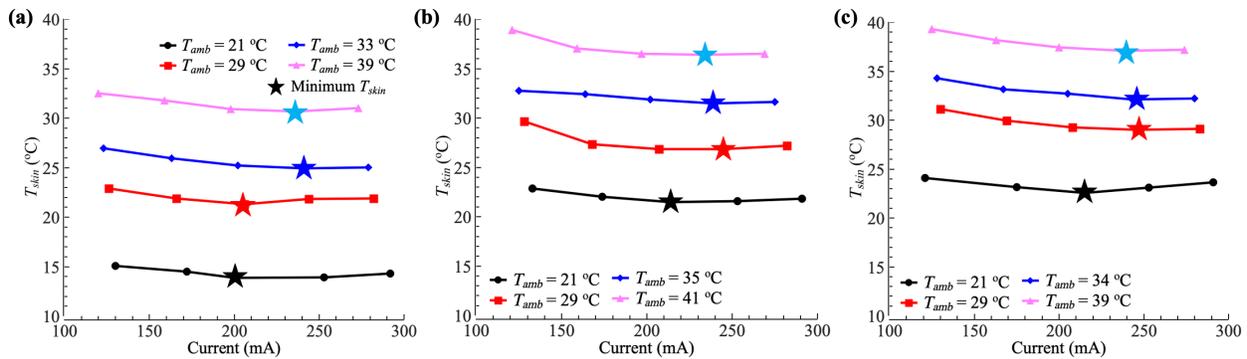

Figure S4. The skin temperature as a function of the applied current at different ambient conditions at three heat loads in single TED: (a) no heat load; (b) $Q_{skin}$ = 80 W m⁻²; (c) $Q_{skin}$ = 96.4 W m⁻²

**Note S5. Long-term Stability of TED-based Cooling Garment**

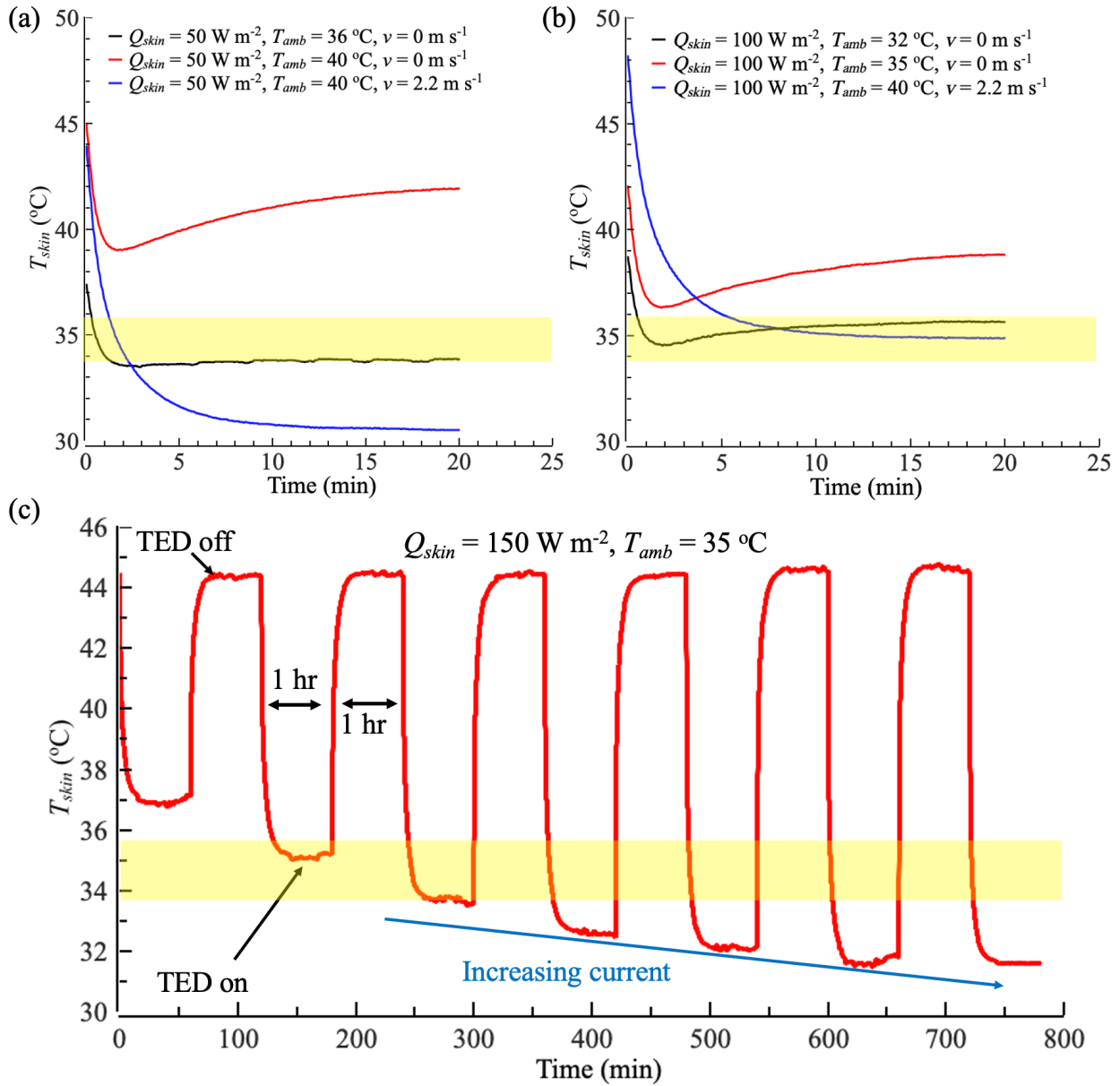

Figure S5. Real-time temperature response at different operation conditions on the manikin model, the shaded area is the thermal comfort zone on the human back region with the temperature range 33.8 – 35.8 °C: (a) 50 W m$^{-2}$ applied heat load; (b) 100 W m$^{-2}$ applied heat load; (c) 150 W m$^{-2}$ applied heat load under cyclic operation conditions at 35 °C ambient temperature with 2.2 m s$^{-1}$ windy condition

**Note S6. TE Materials Impact on Cooling Garments**

In this study, we selected the commercially available TE materials $Bi_2Te_3$ based on our previous studies with an effective ZT value of 0.71[1]. Recently, ZT values range from 0.8 up to 1.48 at room temperature has been experimentally demonstrated in different material systems, such as MgAgSb-based material with a ZT value ~0.8[2], BiSbTe-based material with a ZT value ~1.2[3], and BiTeAg-based material with a ZT value of ~1.48[4]. We evaluated the impact of different ZT values on TED cooling garments with two different aspects (energy efficiency and weight/thickness reduction) via a COMSOL model following our previous work[1].

We began by fixing the same TED design and ambient conditions (ambient temperature, $T_{amb}$=40 °C, under forced convection with a wind speed of 2.2 m/s) while varying the TE materials. To achieve the same cooling power as the current $Bi_2Te_3$-based device, we evaluated the corresponding power consumption. The results, summarized in Figure S6(a), show a significant enhancement in COP, with an approximately 40% improvement achieved by replacing the current TE material with MgAgSb-based material. An additional 30% COP enhancement is predicted with the adoption of TE materials with even higher ZT values.

The slight increase in power consumption observed with BiTeAg-based material can be attributed to its relatively lower Seebeck coefficient (~120 µV/K) compared to BiSbTe-based material (~190 µV/K). Beyond energy savings, the maximum allowable operating ambient temperature can be increased by up to 45%, compared to the current maximum working ambient temperature of 40 °C, under the same heat load of 100 W/m².

By utilizing advanced TE materials, the height and weight of TE pillars can be significantly reduced while maintaining the same cooling power, enabling the development of more compact and lightweight devices. As illustrated in Figure S6(b), a weight reduction of approximately 60% and a thickness reduction of around 40% can be achieved with MgAgSb-based materials. As ZT values increase, thinner TE pillars are sufficient to provide adequate cooling capacity. However, the weight reduction is less pronounced for materials such as BiTeSb and BiTe-Ag systems, which have relatively higher densities (>6000 kg/m³) compared to MgAgSb materials (<5500 kg/m³).

Based on our simulation predictions, MgAgSb-based TE material demonstrates an optimal balance for next-generation TED-based cooling garments, combining superior performance with reduced weight. Further optimization could focus on both the ZT value and material density to achieve even greater performance enhancements.

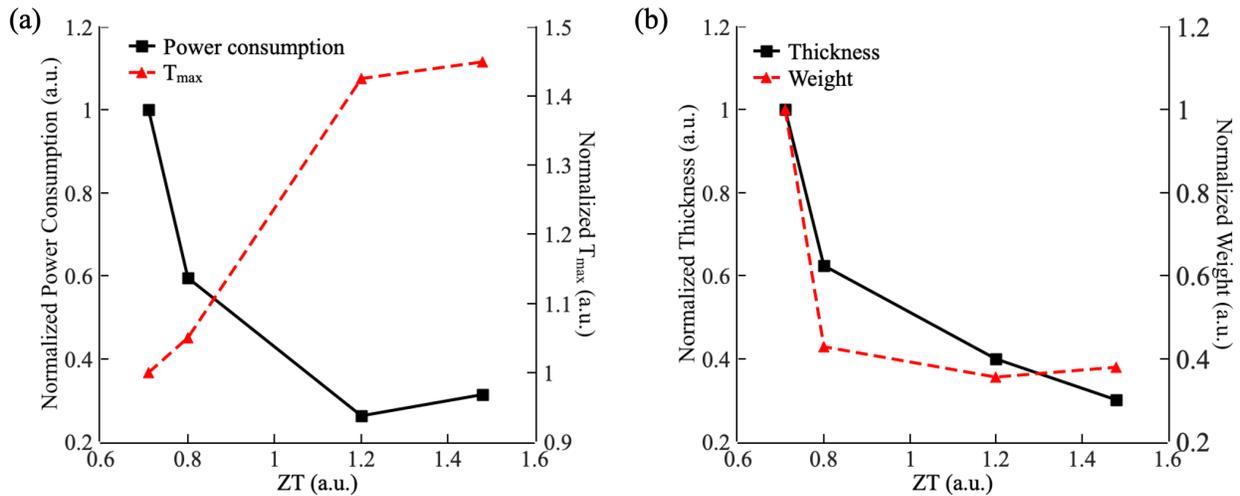

Figure S6. The impact of different TE materials with different ZT value on overall cooling performance: (a) the power consumption and maximum working temperature; (b) the TE pillar thickness and total weight